\documentclass[twoside,fleqn]{article}
\usepackage{espcrc2}
\usepackage{epsf}
\usepackage{amssymb}

\title{How well do domain wall fermions realize chiral symmetry?}

\author{
  George~R.~Fleming
  \thanks{
    In collaboration with
    P.~Chen,
    N.~Christ,
    A.~Kaehler,
    T.~Klassen,
    C.~Malureanu,
    R.~Mawhinney,
    G.~Siegert
    C.~Sui,
    P.~Vranas,
    L.~Wu,
    Y.~Zhestkov.
    Supported in part by DOE grant \# DE-FG02-92ER40699
    and in part by NSF grant \# NSF-PHY96-05199 (PMV).
  }
  \address{Physics Dept., Columbia University, New York NY 10027}
}

\begin{document}

%
%
%
\def\thepage{CU-TP-952}

\begin{abstract}

In the domain wall fermion formulation, chiral symmetry breaking in
full QCD is expected to fall exponentially with the length of the extra
dimension.  We measure the chiral symmetry breaking due to a finite
extra dimension in two ways, which can be affected differently by
finite volume and explicit fermion mass.  For quenched QCD the two
methods generally agree, except for the largest extent of the extra
dimension, which makes the limit uncertain.  We have less data for full
QCD, but see exponential suppression for the method where we have
data.

\end{abstract}

\maketitle

\section{Introduction}
\label{sec:introduction}

Lattice QCD with massless domain wall fermions (including fermion loop
effects) \cite{Kaplan:1992bt}\cite{Furman:1995ky} is expected to have
the ${\rm SU}_L(N_f) \otimes {\rm SU}_R(N_f)$ chiral symmetry of the
continuum when the extent of the extra dimension, $L_s$, becomes
infinite.  For simulations, where the volume is finite and particles
are not strictly massless, reliable techniques are needed to quantify
the symmetry breaking for finite $L_s$.  Such techniques are needed to
see the expected $\exp( -\alpha L_s)$ dependence of chiral breaking for
full QCD and determine if this is also the case for the quenched
theory.

Here we report results from two techniques for measuring chiral
symmetry breaking due to finite $L_s$; the first uses the pion mass and
the second the axial Ward identity.  At zero temperature, the pion mass
is governed by the axial Ward identity.  However, in simulations with
finite volume and with finite quark masses, it is important to check
the agreement between these approaches.

The axial Ward identity is the origin of the Gell-Mann-Oakes-Renner
(GMOR) relation, discussed previously for domain wall fermions in
\cite{Fleming:1998cc}.  The fermion action of \cite{Furman:1995ky} is
used, with the modifications of \cite{Vranas:1997da}. Some details on
the numerical methods are in \cite{Vranas:1998vm}.  See
\cite{Blum:1998ud} for a general review on domain wall fermions and
references.

\section{$m_{\rm res}$ for domain wall fermions}
\label{sec:GMOR}

If the dominant effect of finite $L_s$ is to produce an extra
contribution, $m_{\rm res}$, to the total quark mass, then one would
expect
\begin{equation}
  m_\pi^2 = c_0(V) + c_1(V) * ( m_f + m_{\rm res} ) + \cdots
\end{equation}
where $V$ is the space-time volume and it is expected that $c_0(V)
\rightarrow 0$ as $V \rightarrow \infty$.  For finite volume, a result
we call $m_{\rm res}^{(m_\pi^2)}$, can be found from $m_\pi^2(m_f = 0
)/c_1(V)$, which is $m_{\rm res}$ when $c_0(V) = 0$.

Using the flavor non-singlet axial current in \cite{Furman:1995ky}
we integrate the axial Ward--Takahashi identity to get
\begin{equation}
\label{eq:ward}
\left\langle \bar{q}_0 q_0 \right\rangle
= m_f \chi_\pi + \Delta J_5
\end{equation}
where the pseudoscalar susceptibility is (no sum on $a$)
\begin{equation}
\chi_\pi \equiv 2 \sum_x \left\langle
  \bar{q}_x \gamma_5 \frac{\lambda^a}{2} q_x
  ~ \bar{q}_0 \gamma_5 \frac{\lambda^a}{2} q_0
\right\rangle,
\end{equation}
the additional contribution from chiral mixing due to finite $L_s$ is
\begin{equation}
\Delta J_5 \equiv 2 \sum_x \left\langle
  j_5^a\left(x,L_s/2\right)
  ~ \bar{q}_0 \gamma_5 \frac{\lambda^a}{2} q_0 
\right\rangle,
\end{equation}
and $q_x$ are four-dimensional fermion fields defined from the
appropriate right- and left-handed fields at the boundaries of
the extra dimension.

For large volumes in the chirally broken phase, the pseudoscalar
susceptibility is expected to behave as
\begin{equation}
\label{eq:chi_pi_ansatz}
\chi_\pi = a_{-1}/(m_f+m_{\rm res}) + a_0 + {\cal O}(m_f+m_{\rm res}).
\end{equation}
The first term again says that, for large volumes, the pion is massless
at $m_f = - m_{\rm res}$, while $a_0$ gives the contribution due
to the massive modes.  Clearly, the pion pole contribution only
dominates for large enough volumes and small enough $m_f + m_{\rm
res}$.

$j_5^a\left(x,L_s/2\right)$, a pseudoscalar density located midway
between the domain walls, also has a pole contribution, whose
coefficient is suppressed by propagation from $L_s/2$ to the
boundaries.  Since $\chi_\pi$ and $\Delta J_5$ both have a pole at $m_f
= - m_{\rm res}$ and when the pole terms dominate (\ref{eq:ward})
$\left\langle \bar{q}_0 q_0 \right\rangle$ is finite, we can write in
general
\begin{equation}
\label{eq:chi_pi_j5}
\Delta J_5 =  m_{\rm res}\chi_\pi + b_0 + {\cal O}(m_f+m_{\rm res}).
\end{equation}

We define ${m_{\rm res}}^{\rm(GMOR)}$ by simultaneously fitting
to the form
\begin{equation}
\label{eq:GMOR}
\left\langle \bar{q}_0 q_0 \right\rangle
= (m_f + m_{\rm res}) \chi_\pi + b_0.
\end{equation}
and $\chi_\pi$ as given in (\ref{eq:chi_pi_ansatz}).  For a given
$L_s$, this is a four parameter fit for $a_{-1}, a_0, b_0$ and $m_{\rm
res}^{(GMOR)}$.  Note that only if $b_0/\chi_\pi$ is small, can we get
a reliable estimate for $m_{\rm res}^{(GMOR)}$ from $ \left\langle
\bar{q}_0 q_0 \right\rangle / \chi_\pi - m_f$.  For full QCD, both
$m_{\rm res}^{(GMOR)}$ and $b_0$ should approach zero exponentially in
$L_s$, since both involve propagation from $L_s/2$ to the walls.

\section{$m_{\rm res}$ in quenched QCD}
\label{sec:quench_mres}

\pagenumbering{arabic}
\setcounter{page}{2}

\begin{figure}
  \epsfxsize=\hsize
  \epsfbox{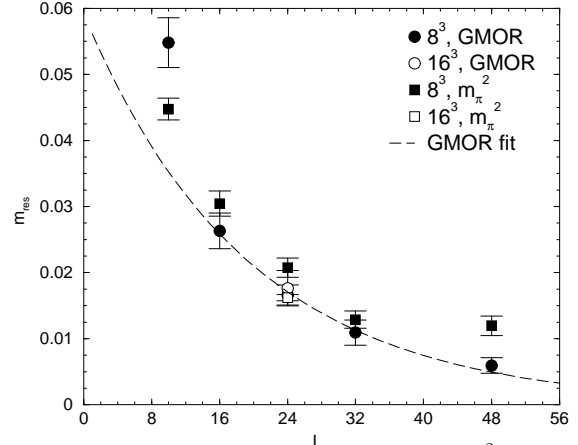}
 \vspace{-1.3cm}
  \caption{
    Comparison of ${m_{\rm res}}^{(m_\pi^2)}$ and ${m_{\rm res}}^{\rm(GMOR)}$
    at quenched $\beta=5.7$
  }
  \label{fig:mres_wn_b5.7}
 \vspace{-0.5cm}
\end{figure}

We first find $m_{\rm res}^{(m_\pi^2)}$ for the the $\beta=5.7$,
$m_0=1.65$, $8^3 \times 32$ quenched domain wall spectrum study we
reported last year \cite{Mawhinney:1998ut}.  For quenched QCD, the
observed zero mode contribution to $\langle \bar{q}_0 q_0 \rangle $ is
small for $m_f \ge 0.02$, so we restrict our attention this mass
range here.  Figure \ref{fig:mres_wn_b5.7}, shows $m_{\rm
res}^{(m_\pi^2)}$ for our $\beta=5.7$, $m_0=1.65$, $8^3 \times 32$
simulations from a correlated, linear fit of $m_\pi^2$ to valence
masses 0.02, 0.06, 0.10, with errors from jacknifing.  The $L_s = 32$
and 48 values are the same within errors, making the large $L_s$ limit
seem non-zero.  For $L_s = 24$ the result for a $16^3 \times 32$
lattice is also shown, revealing that finite volume effects are
noticeable.

Figure \ref{fig:mres_wn_b5.7} also shows $m_{\rm res}^{(GMOR)}$ from a
correlated fit to (\ref{eq:GMOR}) and (\ref{eq:chi_pi_ansatz}).  The
$16^3$ GMOR point is on top of the $8^3$ point in the plot.
The fits $m_{\rm res} = 0.059(14) \exp[-0.052(10) L_s]$ and
$b_0 = -0.0035(11) \exp[ -0.051(13) L_s ]$ (not shown)
are consistent with $\Delta J_5$ vanishing in the $L_s\to\infty$ limit.

The agreement between the two methods for $L_s \le 32$ is reasonable
and only occurs since $b_0$ is included in the fits.  $m_{\rm
res}^{(GMOR)}$ is volume independent for $L_s = 24$, while $m_{\rm
res}^{(m_\pi^2)}$ is not.  The discrepancy for $L_s = 48$ may be due to
finite volume, but needs further study.

\section{$m_{\rm res}$ in $N_f=2$ QCD}
\label{sec:dyn_mres}

\begin{figure}
  \epsfxsize=\hsize
  \epsfbox{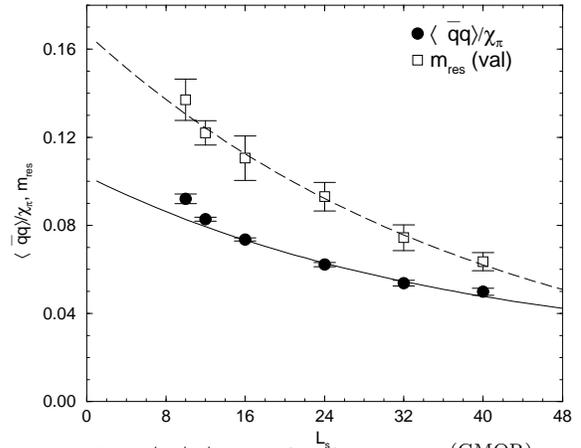}
  \vspace{-1.3cm}
  \caption{
    $\left\langle\bar{q}q\right\rangle / \chi_\pi$
    and valence ${m_{\rm res}}^{\rm(GMOR)}$
    on $8^3\times 4, \beta=5.2, m_0=1.9, m_f=0.02$ lattices.
  }
  \vspace{-0.5cm}
  \label{fig:mres_wd_8nt4_b5.2_h1.9_m0.02}
\end{figure}

For full QCD, we have done extensive simulations with the Wilson gauge
action and domain wall fermions on $8^3\times 4$ volumes for several
values of $L_s$ with $\beta=5.2$, $m_0=1.9$ and $m_f=0.02$.  For these
lattices, which are in the low temperature phase, we show
$\left\langle\bar{q}q\right\rangle / \chi_\pi$ in Figure
\ref{fig:mres_wd_8nt4_b5.2_h1.9_m0.02}.  An exponential fit, yielding
$\left\langle\bar{q}q\right\rangle / \chi_\pi = 0.02 +
  0.082(3) \exp[-0.027(2) L_s]$ for $16 \le L_s \le 40$ with
$\chi^2/N_{\rm dof} = 2.76/2$, is also shown.

We have also done uncorrelated fits, using valence masses between 0.02
and 0.14, to extract $m_{\rm res}^{(GMOR)}$ and $b_0$, since for the
dynamical simulations there is not enough data to resolve the
covariance matrix.  (For the quenched case there was little difference
between the correlated and uncorrelated fits.)  All fits have $N_{\rm
dof}=4$ and $\chi^2/N_{\rm dof} \lesssim 1$.  ${m_{\rm
res}}^{\rm(GMOR)}$ is also shown in Figure
\ref{fig:mres_wd_8nt4_b5.2_h1.9_m0.02} and the dashed line fit is
${m_{\rm res}}^{\rm(GMOR)} = 0.17(2) \exp[-0.026(6) L_s]$ for $10 \le
L_s \le 40$ with $\chi^2/N_{\rm dof} = 0.35/4$.  $b_0$ is not shown,
but also fits the exponential form $b_0 = -0.0100(16) \exp[-0.0147(67)
L_s]$ with $\chi^2/N_{\rm dof} = 0.20/4$ over the same range in $L_s$.

\begin{table}
  \centering
  \caption{
    Valence $m_{\rm res}$ comparison on $8^3\times 32$ lattices
  }
  \label{tab:mres_val_8nt32}
  \begin{tabular}{rllll}
    \multicolumn{1}{c}{$L_s$} &
    \multicolumn{1}{c}{$m_f$} &
    \multicolumn{1}{c}{$m_{\rm res}^{(m_\pi^2)}$} &
    \multicolumn{1}{c}{$m_{\rm res}^{\rm(GMOR)}$} &
    \multicolumn{1}{c}{$-b_0$} \\
    \hline
    \multicolumn{5}{|c|}{Wilson gauge action, $\beta=5.325, m_0=1.9$} \\
    \hline
    24 & 0.02 & 0.0622(9)  & 0.058(2) & 0.0047(3) \\
       & 0.06 & 0.0645(6)  & 0.059(2) & 0.0046(2) \\
    \hline
    \multicolumn{5}{|c|}{Iwasaki gauge action,  $\beta=1.9, m_0=1.9$} \\
    \hline
    24 & 0.02 & 0.0401(5)  & 0.038(2) & 0.0028(2) \\
    \hline
    \multicolumn{5}{|c|}{Iwasaki gauge action, $\beta=2.0, m_0=1.9$} \\
    \hline
    24 & 0.02 & 0.0158(9)  & 0.015(2) & 0.0013(3) \\
       & 0.06 & 0.019(1)   & 0.017(3) & 0.0015(4) \\
    48 & 0.02 & 0.0073(16) & 0.011(2) & 0.0010(2) \\
  \end{tabular}
  \vspace{-0.8cm}
\end{table}

In Table \ref{tab:mres_val_8nt32} we compare the two methods of
extracting $m_{\rm res}$ using valence spectrum data from 
$N_f=2, 8^3\times 32$ scale setting calculations \cite{Wu:1999}.
All data was fit for $0.02 \le m_f^{\rm(val)} \le 0.1$.  For the
GMOR fit, $N_{\rm dof}=2$ and $\chi^2/N_{\rm dof} \lesssim 1$ for
all fits.  We note that the two methods agree within statistics.

\begin{table}
  \centering
  \caption{
    Dynamical $m_{\rm res}$ comparison on $8^3\times 32$ lattices
  }
  \label{tab:mres_dyn_8nt32}
  \begin{tabular}{llll}
    \multicolumn{1}{c}{$\beta$} &
    \multicolumn{1}{c}{$m_{\rm res}^{(m_\pi^2)}$} &
    \multicolumn{1}{c}{$m_{\rm res}^{\rm(GMOR)}$} &
    \multicolumn{1}{c}{$-b_0$} \\
    \hline
    \multicolumn{4}{|c|}{$m_0=1.9, L_s=24$} \\
    \hline
    5.325 & 0.059(2) & 0.053(7) & 0.004(1) \\
    2.0   & 0.013(2) & 0.014(5) & 0.0011(9) \\
  \end{tabular}
  \vspace{-0.7cm}
\end{table}

We can also calculate $m_{\rm res}$ both ways but with data as a
function of the dynamical mass.  Since there are only two dynamical
masses, both methods are unconstrained so the errors quoted come from
naive extrapolation.  The results are summarized in Table
\ref{tab:mres_dyn_8nt32}.

\section{Conclusions}
\label{sec:conclusions}

We have gotten good agreement between $m_{\rm res}^{(m_\pi^2)}$ and
$m_{\rm res}^{(GMOR)}$ for a wide range of quenched and dynamical
simulations by including the non-pole terms in the susceptibilities.
From our current data, $m_{\rm res}^{(GMOR)}$ appears less volume
dependent.  For the quenched simulations at $L_s = 48$, the two methods
do not agree, possibly as a result of finite volume effects.  This case
warrants further study.

Whether chiral symmetry is fully restored for quenched simulations in
the $L_s\to\infty$ limit of domain wall fermions is still an open
question.  For $N_f=2$ QCD, $m_{\rm res}^{(GMOR)}$ falls exponentially,
even at quite strong coupling. The rate of chiral symmetry restoration
is very slow leaving much room for improvement.

All numerical calculations were performed on the 400 Gflops QCDSP
machine at Columbia and the 6 Gflops QCDSP machine at Ohio State.


\begin{thebibliography}{99}

\bibitem{Kaplan:1992bt}
D.B.~Kaplan,
Phys. Lett. {\bf B288}, 342 (1992)
.
\bibitem{Furman:1995ky}
V.~Furman and Y.~Shamir,
Nucl. Phys. {\bf B439}, 54 (1995)
.

\bibitem{Fleming:1998cc}
G.R.~Fleming {\it et al.},
Nucl.\ Phys.\ Proc.\ Suppl.\ {\bf 73}, 207 (1999)
.


\bibitem{Vranas:1997da}
P.M.~Vranas,
Phys.\ Rev.\ {\bf D57}, 1415 (1998)
.

\bibitem{Vranas:1998vm}
P.M.~Vranas {\it et al.},
Nucl.\ Phys.\ Proc.\ Suppl.\ {\bf 73}, 456 (1999)
.

\bibitem{Blum:1998ud}
T.~Blum,
Nucl.\ Phys.\ Proc.\ Suppl.\ {\bf 73}, 167 (1999)
.

\bibitem{Mawhinney:1998ut}
R.~Mawhinney {\it et al.},
Nucl.\ Phys.\ Proc.\ Suppl.\ {\bf 73}, 204 (1999)
.

\bibitem{Wu:1999}
L.~Wu,
these proceedings
.

\bibitem{Vranas:1999}
P.M.~Vranas,
these proceedings
.

\end{thebibliography}
\end{document}